\documentclass[10pt,a4paper,twocolumn]{article}

\usepackage[utf8]{inputenc}
\usepackage[T1]{fontenc}
\usepackage{lmodern}
\usepackage[british]{babel}
\usepackage{microtype}
\usepackage{geometry}
\usepackage{setspace}
\usepackage{titlesec}
\usepackage{abstract}
\usepackage{booktabs}
\usepackage{tabularx}
\usepackage{multirow}
\usepackage{array}
\usepackage{longtable}
\usepackage{xcolor}
\usepackage{enumitem}
\usepackage{hyperref}
\hypersetup{
    colorlinks=true,
    linkcolor=black,
    citecolor=black,
    urlcolor=blue,
    pdftitle={The AI Resilience Gap},
    pdfauthor={Jonathan Shelby},
    pdfkeywords={artificial intelligence, operational resilience, DORA, EU AI Act, critical third parties, concentration risk, financial services}
}
\usepackage{graphicx}
\usepackage{tikz}
\usetikzlibrary{shapes.geometric, arrows.meta, positioning, fit, calc, backgrounds}
\usepackage{footnote}
\usepackage{url}
\usepackage[numbers,sort&compress]{natbib}
\usepackage{xspace}

\newcommand{\CSM}{\textsc{csm}\xspace}
\newcommand{\ARF}{\textsc{arf}\xspace}

\newcolumntype{L}[1]{>{\raggedright\arraybackslash}p{#1}}
\newcolumntype{C}[1]{>{\centering\arraybackslash}p{#1}}

\titleformat{\section}{\large\bfseries}{\thesection.}{0.5em}{}
\titleformat{\subsection}{\normalsize\bfseries}{\thesubsection.}{0.5em}{}
\titleformat{\subsubsection}{\normalsize\itshape}{\thesubsubsection.}{0.5em}{}

\tikzset{
    proc/.style={rectangle, rounded corners, draw=black, fill=blue!5, thick, text width=3.1cm, align=center, minimum height=1.0cm, font=\small},
    dec/.style={diamond, aspect=2, draw=black, fill=orange!10, thick, text width=2.3cm, align=center, inner sep=1pt, font=\small},
    term/.style={rectangle, draw=black, fill=gray!8, thick, text width=3.1cm, align=center, minimum height=0.9cm, font=\small},
    flow/.style={-{Stealth[length=2.5mm]}, thick},
    quad/.style={rectangle, draw=black, thick, text width=4.6cm, align=center, minimum height=2.6cm, font=\small}
}

\begin{document}

\title{\vspace{-1.5cm}\textbf{The AI Resilience Gap: Bringing Artificial\\ Intelligence Inside the Operational Resilience Perimeter}}

\author{
    Jonathan Shelby\\
    \small Department of Computer Science, University of Oxford\\
    \small Hertford College, Oxford, United Kingdom\\
}
\date{\small July 2026}

\twocolumn[
\begin{@twocolumnfalse}
\maketitle
\begin{onecolabstract}
\noindent The rapid adoption of artificial intelligence across regulated firms has produced an extensive governance response oriented around \emph{trustworthiness}: the EU AI Act, ISO/IEC 42001, the NIST AI Risk Management Framework, and the United Kingdom's principles-based approach all address safety, fairness, transparency, and model risk. That response is necessary but incomplete. It does not, on its own, address \emph{operational resilience}: the continuity of important business services under severe but plausible disruption, the substitutability of AI components, and the concentration of dependency on the small number of firms that supply frontier models. This paper argues that AI adoption creates a resilience obligation that is distinct from, and inadequately covered by, the trustworthy-AI stack, and that United Kingdom financial authorities are already closing this gap through the Financial Policy Committee's systemic analysis, the Critical Third Parties regime, and the May 2026 joint statement on frontier AI and cyber resilience. We map the two regulatory logics, identify the structural gap between them, and propose the AI Resilience Framework: a regime-agnostic method for bringing AI dependencies inside the operational resilience perimeter through dependency mapping, a criticality-substitutability tiering, the extension of impact tolerances to AI-specific failure modes, an explicit fallback doctrine, and provider-level concentration management. The framework gives chief information security officers, security architects, and boards an actionable route from AI governance policy to demonstrable resilience. This work extends a companion analysis of the United Kingdom cyber resilience regulatory stack \cite{shelby2026regstack} into the artificial intelligence dimension.\\

\noindent \textbf{Keywords:} artificial intelligence, operational resilience, EU AI Act, DORA, Critical Third Parties, concentration risk, impact tolerance, model risk, financial services
\end{onecolabstract}
\vspace{0.6cm}
\end{@twocolumnfalse}
]

\section{Introduction}

Artificial intelligence has moved from pilot to production across United Kingdom financial services. The third Bank of England and Financial Conduct Authority survey of AI in the sector, published in November 2024, reported that around three quarters of firms were already using AI, with a further tenth planning adoption within three years, and that foundation models, including large language models, already accounted for roughly a sixth of use cases \cite{boefcaaisurvey2024}. A majority of use cases involved some degree of autonomous decision-making, even though only a small fraction were described as fully autonomous. The direction of travel is unambiguous. AI is no longer confined to marketing analytics and back-office productivity; it is entering credit decisioning, transaction monitoring, trading, and the operation of customer-facing services that firms are obliged to keep running.

The regulatory and standards response to this shift has been substantial. At the European level, the AI Act establishes a horizontal, risk-tiered regime for AI systems and general-purpose models \cite{euaiact2024}. Internationally, ISO/IEC 42001 provides a certifiable management-system standard for AI \cite{iso42001}, and the NIST AI Risk Management Framework offers a voluntary structure for identifying and managing AI risks, extended in 2024 by a profile for generative AI \cite{nistairmf2023, nistgenai2024}. In the United Kingdom, financial regulators have chosen a principles-based, technology-neutral path, declining to introduce AI-specific rules and instead relying on existing frameworks such as the Consumer Duty, the Senior Managers and Certification Regime, model risk management expectations, and operational resilience requirements \cite{boefpcai2025, treasurycommai2026}.

These instruments share an orientation. They are concerned, first and foremost, with whether an AI system is trustworthy: whether it is safe, fair, explainable, well-documented, adequately governed, and free from prohibited or high-risk misuse. This is the right set of questions, and it is not the only set. A separate body of regulation asks a different question entirely. The operational resilience regime operated by the Financial Conduct Authority and the Prudential Regulation Authority requires firms to identify their important business services, set impact tolerances for the maximum tolerable disruption to each, and remain within those tolerances under severe but plausible scenarios \cite{fcaopres2021, praopres2021}. The Digital Operational Resilience Act imposes a parallel discipline across the European Union \cite{dora2022}, and the United Kingdom Critical Third Parties regime extends supervisory attention to the concentrated providers on which the sector collectively depends \cite{fsma2023, ss624ctp2024}. This body of regulation is not concerned with whether a system is trustworthy. It is concerned with whether the service survives when something goes wrong.

The central argument of this paper is that these two logics have not been joined, and that the gap between them is where the material risk of AI adoption now sits. A firm can deploy an AI system that is entirely trustworthy in the sense the AI Act intends, accurate, fair, documented, and overseen, and that is simultaneously a severe resilience liability: a single, unsubstitutable dependency on one external model provider, embedded in an important business service, with no rehearsed path to continue operating if that provider degrades, changes its model, or becomes unavailable. Trustworthiness and resilience are different properties. Managing the first does not deliver the second.

This gap is not hypothetical, and United Kingdom authorities have begun to close it. In April 2025 the Financial Policy Committee published an assessment of AI in the financial system that identified concentration in third-party model and infrastructure provision as a potential source of systemic risk \cite{boefpcai2025}. The Financial Stability Board reached similar conclusions internationally \cite{fsbai2024, fsbaiadoption2025}. In January 2026 the Financial Conduct Authority launched a long-term review of AI in retail financial services that explicitly raised the sector's dependence on a small number of dominant AI infrastructure providers \cite{fcamillsreview2026}. And in May 2026 the Bank of England, the Financial Conduct Authority, and HM Treasury issued a joint statement on frontier AI and cyber resilience, reinforcing existing expectations on governance, vulnerability management, third-party oversight, and recovery in the face of AI-enabled attacks \cite{boefcahmtfrontier2026}. The supervisory perimeter is moving toward AI resilience. Firms whose internal operating models still treat AI purely as a model-risk and ethics question are preparing for the wrong examination.

This paper makes four contributions:

\begin{enumerate}[leftmargin=1.4cm, itemsep=2pt]
    \item It distinguishes the \emph{trustworthy-AI} regulatory logic from the \emph{operational-resilience} logic and shows, through a structured comparison, why the former does not discharge the obligations of the latter.
    \item It characterises the distinctive failure modes of AI systems, in particular silent degradation and non-determinism, that the operational resilience regime was not designed to detect, and it situates foundation-model provider concentration as the AI analogue of cloud concentration risk.
    \item It proposes the AI Resilience Framework (\ARF), whose core is a Criticality-Substitutability Matrix (\CSM) for classifying AI dependencies, together with an explicit fallback doctrine and a provider-level concentration treatment.
    \item It translates the framework into practitioner guidance for security leaders, architects, model-risk functions, boards, and compliance teams, and maps it onto the United Kingdom and European instruments now in force.
\end{enumerate}

The paper positions itself within a companion body of work. A practitioner analysis of the United Kingdom Cyber Security and Resilience Bill \cite{shelby2026csr} argued that expanded incident-reporting and supply-chain duties act as an architectural forcing function. A subsequent analysis of the wider United Kingdom cyber resilience regulatory stack \cite{shelby2026regstack} developed a Compliance Convergence Framework for satisfying multiple overlapping regimes from a single control base. The present paper extends that convergence argument into the AI dimension, where the overlap is not between two resilience regimes but between the resilience regime and the emerging trustworthy-AI regime.

The remainder of the paper is organised as follows. Section 2 sets out the two regulatory logics and the structural gap between them. Section 3 argues that AI is properly understood as an operational-resilience problem and examines its distinctive failure modes and concentration dynamics. Section 4 presents the AI Resilience Framework. Section 5 applies the framework to three illustrative dependencies. Section 6 draws out the implications for practitioners, and Section 7 concludes.

\section{Two Regulatory Logics}

The instruments that govern AI in a regulated firm fall into two families with different objectives, different units of analysis, and different conceptions of what it means for a system to fail. Understanding the distinction is the precondition for seeing the gap.

\subsection{The Trustworthy-AI Stack}

The dominant framework is the European Union AI Act, which entered into force on 1 August 2024 and applies its obligations in phases \cite{euaiact2024}. Prohibited practices and AI literacy obligations took effect in February 2025, and the obligations for providers of general-purpose AI models took effect in August 2025. The most demanding provisions, those for high-risk systems, were originally scheduled for August 2026 and August 2027, but the Digital Omnibus on AI, on which a provisional political agreement was reached on 7 May 2026 and which remains subject to formal adoption, deferred the principal high-risk obligations for use-based Annex III systems to December 2027 and for product-embedded Annex I systems to August 2028 \cite{euaiomnibus2026}. The Act is extraterritorial in effect: a United Kingdom firm whose AI systems are placed on the European Union market, or whose outputs affect users in the Union, may be in scope regardless of where it is established. Its logic is one of product safety and fundamental-rights protection. It classifies systems by the risk they pose to individuals and society, and it imposes documentation, data governance, human oversight, transparency, and conformity obligations calibrated to that risk.

Alongside the Act sit two voluntary but influential instruments. ISO/IEC 42001, published in 2023, defines a certifiable AI management system in the familiar plan-do-check-act structure, allowing firms to demonstrate governance maturity in a form auditors recognise \cite{iso42001}. The NIST AI Risk Management Framework, published in 2023 and supplemented in 2024 by a generative-AI profile, organises AI risk management around the functions of governing, mapping, measuring, and managing, and has become a common reference point for firms operating on both sides of the Atlantic \cite{nistairmf2023, nistgenai2024}.

The United Kingdom has deliberately not legislated an equivalent horizontal regime for financial services. The Financial Conduct Authority has stated repeatedly that it does not intend to introduce AI-specific rules, on the grounds that a technology evolving on a timescale of months is ill-suited to prescriptive regulation, and that it will instead supervise AI through existing outcomes-focused frameworks \cite{treasurycommai2026}. The Prudential Regulation Authority's model risk management expectations for banks, set out in supervisory statement SS1/23, are technology-agnostic but were deliberately drafted to capture factors relevant to AI models \cite{ss123mrm2023}. This principles-based posture is coherent, but it places the burden on firms to work out how general rules apply to specific AI deployments.

What unites this family is its objective. Whether through binding law, certifiable standard, or supervisory principle, the trustworthy-AI stack asks whether an AI system is fit to be relied upon: whether its outputs are accurate and fair, whether its operation is transparent and accountable, and whether its risks to individuals have been identified and mitigated. These are questions about the quality of the system.

\subsection{The Operational-Resilience Stack}

The second family begins from a different premise. It assumes that systems will fail, and it asks whether the services that depend on them can continue, or recover quickly enough, when they do. In the United Kingdom, the operational resilience regime introduced by the Financial Conduct Authority and the Prudential Regulation Authority requires firms to identify their important business services, set an impact tolerance defining the maximum tolerable duration or extent of disruption to each, map the resources on which each service depends, and demonstrate that they can remain within tolerance under severe but plausible scenarios \cite{fcaopres2021, praopres2021}. The transitional period for this regime ended and firms moved to steady-state expectations during 2025, so the obligation to remain within impact tolerances is now live rather than prospective.

At the European level, the Digital Operational Resilience Act imposes a comprehensive information and communication technology risk-management framework, including incident classification and reporting, digital operational resilience testing, and third-party risk management, with an oversight regime for critical ICT third-party providers \cite{dora2022}. The United Kingdom has constructed an analogous mechanism for concentration risk. The Financial Services and Markets Act 2023 created a regime under which HM Treasury may designate critical third parties whose failure would threaten the stability of, or confidence in, the United Kingdom financial system, and the Bank, the Prudential Regulation Authority, and the Financial Conduct Authority jointly published rules for that regime, in supervisory statement SS6/24, in November 2024 \cite{fsma2023, ss624ctp2024}. Commentary anticipates that major AI and cloud providers will be among the entities designated under this regime \cite{boefpcai2025}.

What unites this family is a concern not with the quality of a system but with the continuity of a service and the substitutability of its components. Its central artefacts, the important business service, the impact tolerance, the substitution assessment, and the concentration analysis, are indifferent to whether a dependency is an AI model, a database, or a payment rail. They ask only what happens to the service if the dependency is lost.

The two stacks are also proceeding on different timetables, which conditions how firms should sequence their response. Table~\ref{tab:timeline} sets out the principal milestones now in force or scheduled. The operational resilience obligations are already live: the United Kingdom regime reached steady state during 2025, and the Digital Operational Resilience Act has applied since January 2025. The trustworthy-AI obligations are arriving more gradually, with the most demanding high-risk provisions of the EU AI Act deferred, under the Digital Omnibus agreed in May 2026, to late 2027 and 2028. The practical implication is that the resilience obligations bite first. A firm cannot defer its AI resilience work to align with the AI Act high-risk timetable, because the operational resilience regime already requires it.

\begin{table*}[tp]
\centering
\caption{Principal milestones across the two stacks (selected).}
\label{tab:timeline}
\small
\begin{tabular}{L{2.6cm} L{6.2cm} L{4.4cm}}
\toprule
\textbf{Date} & \textbf{Milestone} & \textbf{Stack} \\
\midrule
Nov 2024 & Bank and FCA third AI survey; joint Critical Third Parties rules (SS6/24) & Resilience \\
\addlinespace
Jan 2025 & Digital Operational Resilience Act applies & Resilience \\
\addlinespace
Feb 2025 & EU AI Act prohibited practices and AI literacy obligations apply & Trustworthy-AI \\
\addlinespace
Apr 2025 & Financial Policy Committee report on AI in the financial system & Resilience \\
\addlinespace
During 2025 & United Kingdom operational resilience regime reaches steady state & Resilience \\
\addlinespace
Aug 2025 & EU AI Act general-purpose AI model obligations apply & Trustworthy-AI \\
\addlinespace
Jan 2026 & FCA long-term review of AI in retail financial services (Mills Review) & Resilience \\
\addlinespace
May 2026 & Bank, FCA and HM Treasury joint statement on frontier AI and cyber resilience; Digital Omnibus provisional agreement & Both \\
\addlinespace
Aug 2026 & EU AI Act transparency obligations apply & Trustworthy-AI \\
\addlinespace
Dec 2027 & EU AI Act high-risk (Annex III) obligations apply, as deferred by the Omnibus & Trustworthy-AI \\
\bottomrule
\end{tabular}
\end{table*}

\subsection{The Gap Between Them}

The two logics are complementary in principle and disconnected in practice. Table~\ref{tab:logics} sets out the contrast across six dimensions. The trustworthy-AI stack takes the \emph{model} or \emph{system} as its unit of analysis; the resilience stack takes the \emph{service}. The trustworthy-AI stack conceives failure as harm, bias, or opacity; the resilience stack conceives failure as unavailability or intolerable disruption. The trustworthy-AI stack treats third parties primarily as sources of documentation and conformity obligations; the resilience stack treats them as concentration risks and points of systemic fragility. A firm can satisfy every requirement in the first column and remain wholly exposed in the second.

\begin{table*}[tp]
\centering
\caption{Two regulatory logics for artificial intelligence in regulated firms.}
\label{tab:logics}
\small
\begin{tabular}{L{3.0cm} L{5.4cm} L{5.4cm}}
\toprule
\textbf{Dimension} & \textbf{Trustworthy-AI stack} & \textbf{Operational-resilience stack} \\
\midrule
Primary objective & System is safe, fair, transparent, accountable & Service continues or recovers under disruption \\
\addlinespace
Unit of analysis & The model or AI system & The important business service \\
\addlinespace
Concept of failure & Harm, bias, opacity, prohibited use & Unavailability, intolerable disruption, non-recovery \\
\addlinespace
Treatment of third parties & Documentation, conformity, provider obligations & Concentration, substitutability, systemic fragility \\
\addlinespace
Principal instruments & EU AI Act; ISO/IEC 42001; NIST AI RMF; UK principles & UK operational resilience; DORA; UK Critical Third Parties \\
\addlinespace
Governing question & Is the system fit to be relied upon? & Does the service survive when it is not? \\
\bottomrule
\end{tabular}
\end{table*}

The disconnection is reinforced by how firms are organised. Trustworthy-AI obligations tend to be owned by model-risk functions, data-science leadership, and newly created AI governance committees. Resilience obligations are owned by operational resilience teams, chief operating officers, and the senior managers accountable under the Senior Managers and Certification Regime. These functions rarely share a register. The result is that AI dependencies are frequently assessed for model risk and fairness without ever being mapped into an important business service, assigned an impact tolerance, or subjected to a substitution assessment. The dependency is governed, but it is not made resilient.

\section{AI as an Operational-Resilience Problem}

If the gap is to be closed, AI must be treated as a first-class subject of the resilience regime rather than an afterthought to model governance. This section establishes why that treatment is warranted and identifies the properties of AI systems that make it non-trivial.

\subsection{AI Systems Are In Scope of the Resilience Regime}

There is no ambiguity of scope. Under the Digital Operational Resilience Act, an AI system that supports a financial service is information and communication technology, and it inherits the full weight of the ICT risk-management framework, including resilience testing, incident reporting, and third-party oversight \cite{dora2022}. Under the United Kingdom regime, an AI system that supports an important business service is a resource on which that service depends, and it must be mapped, tolerance-assessed, and included in scenario testing \cite{fcaopres2021}. The Financial Conduct Authority has been explicit that its operational resilience rules apply to AI as they apply to any other technology \cite{treasurycommai2026}. The question is therefore not whether AI is in scope, but whether firms are actually bringing it inside the perimeter. The evidence of siloed governance suggests that many are not.

\subsection{Distinctive Failure Modes}

The reason this matters is that AI systems fail in ways the resilience regime was not designed to detect. Conventional resilience thinking is oriented toward discrete, observable failure: a service is available or it is not, a system is up or it is down, and an impact tolerance is breached when the outage exceeds a defined duration. AI systems introduce failure modes that do not fit this binary.

The most important is silent degradation. A model whose performance drifts, because the data distribution it faces has shifted, or because an upstream provider has changed the underlying model, does not stop responding. It continues to return outputs, on time and in the expected format, that are progressively less correct. Monitoring built to detect unavailability will report the service as healthy while its decisions deteriorate. This is a resilience failure that presents as full availability, and it is invisible to controls calibrated for outages. The analogous problem in distributed systems is well known as grey failure, in which a component is neither fully working nor fully failed; AI makes grey failure the normal case rather than the exception.

To make the problem concrete, consider a credit-decisioning model whose input population shifts gradually as a firm enters a new customer segment. The model continues to score every application within its latency budget, and every availability metric remains green, yet its decisions become progressively miscalibrated for the new population. Under a conventional duration-based impact tolerance, no breach is recorded, because the service never went down. The harm accrues in the space the resilience regime was not built to observe: the service was continuously available and continuously wrong. Detecting this failure requires monitoring the distribution of inputs and the calibration of outputs, not the uptime of the endpoint, and defining in advance the level of miscalibration at which the service is deemed to have failed.

Related failure modes compound the difficulty. Non-determinism means that identical inputs may not produce identical outputs, which frustrates the reproducibility that testing and incident investigation assume. Adversarial input, including prompt injection against systems that combine untrusted input with the ability to act, converts an integrity problem into an availability and safety problem. And where an AI component has been granted autonomy, its errors propagate at machine speed and machine scale, enlarging the blast radius of a failure before a human can intervene. None of these modes is captured by a simple duration-based impact tolerance. A resilience regime that measures only whether the service responded, and not whether it responded correctly, will systematically under-detect AI failure.

\subsection{Concentration as Systemic Fragility}

The second dimension on which AI stresses the resilience regime is concentration. The capability that makes foundation models attractive is supplied by a small number of firms, and the infrastructure on which those firms depend is more concentrated still. A sector that builds many services on the same handful of external models acquires a shared dependency that individual firms cannot see from their own vantage point and cannot mitigate through their own controls. This is the AI analogue of the cloud-concentration problem that motivated the Critical Third Parties regime, and regulators have made the parallel explicit. The Financial Policy Committee identified concentration in third-party model and infrastructure provision as a channel through which AI adoption could pose systemic risk \cite{boefpcai2025}, the Treasury Committee heard evidence that United Kingdom firms are over-reliant on a small number of providers \cite{treasurycommai2026}, and the AI Consortium established by the Bank in 2025 took concentration among third-party model providers as a primary workstream \cite{boefpcai2025}.

Concentration converts individually reasonable decisions into a collective vulnerability. If a widely used model exhibits a correlated failure, a systematic bias, an availability incident, or an exploited vulnerability, the effect is felt simultaneously across every firm that relies on it, in a way that resembles a monoculture more than a diversified market. The substitution assessment that the resilience regime demands must therefore be performed not only at the level of the individual firm but with an awareness of sector-level correlation that no single firm can fully observe. This is precisely the circumstance in which the Critical Third Parties regime is intended to operate, and it is why the anticipated designation of major AI and cloud providers matters for every firm downstream of them.

\subsection{AI as a Threat as Well as a Dependency}

Finally, AI stresses resilience from the attacker's side. The joint statement issued by the Bank of England, the Financial Conduct Authority, and HM Treasury in May 2026 warned that the cyber capabilities of frontier AI models already exceed what a skilled human practitioner can achieve, operating at greater speed, scale, and lower cost, and it called on firms to strengthen vulnerability management, third-party oversight, and recovery capability accordingly \cite{boefcahmtfrontier2026}. The statement introduced no new rules, but it reframed frontier AI as a resilience threat and not merely a governance subject. A complete treatment of AI resilience must therefore address both the AI a firm depends upon and the AI that may be used against it. The framework that follows concentrates on the former, which is the less-developed half of current practice, while noting where the two connect.

\section{The AI Resilience Framework}

The AI Resilience Framework (\ARF) is a method for bringing AI dependencies inside the operational resilience perimeter using the machinery firms already possess. Its guiding principle, consistent with the convergence argument developed for the wider regulatory stack \cite{shelby2026regstack}, is deduplication: rather than constructing a parallel AI-risk silo, firms should extend the important business service and impact tolerance constructs they already maintain to encompass their AI dependencies. The framework has five steps.

\subsection{Step One: Map AI Dependencies to Important Business Services}

The first step is to make the dependency visible. For each important business service, the firm identifies every AI component on which that service relies, whether developed internally, embedded in a purchased product, or consumed as an external model service. This mapping is frequently incomplete in practice, because AI is often introduced below the level at which services are catalogued, embedded in a vendor product or wired into a workflow by a delivery team without a corresponding entry in the resilience register. Until the dependency appears on the map, none of the subsequent steps can occur. The output of this step is a register of AI dependencies indexed by the services they support.

\subsection{Step Two: The Criticality-Substitutability Matrix}

The core of the framework is a classification of each mapped dependency along two axes. The first axis is \emph{criticality}: the contribution the AI component makes to an important business service and, correspondingly, the severity with which its failure would erode the service's impact tolerance. The second axis is \emph{substitutability}: the ease with which the firm could continue to deliver the service if the AI component were lost or degraded. We distinguish three substitutability classes. A dependency is \emph{substitutable} if an alternative model or provider could be adopted within acceptable effort and time. It is \emph{degradable} if the service could revert to a reduced or non-AI path, such as a manual process or a deterministic rules engine, within impact tolerance. It is \emph{irreducible} if there is no viable fallback: the AI component is load-bearing and unique, and its loss takes the service with it.

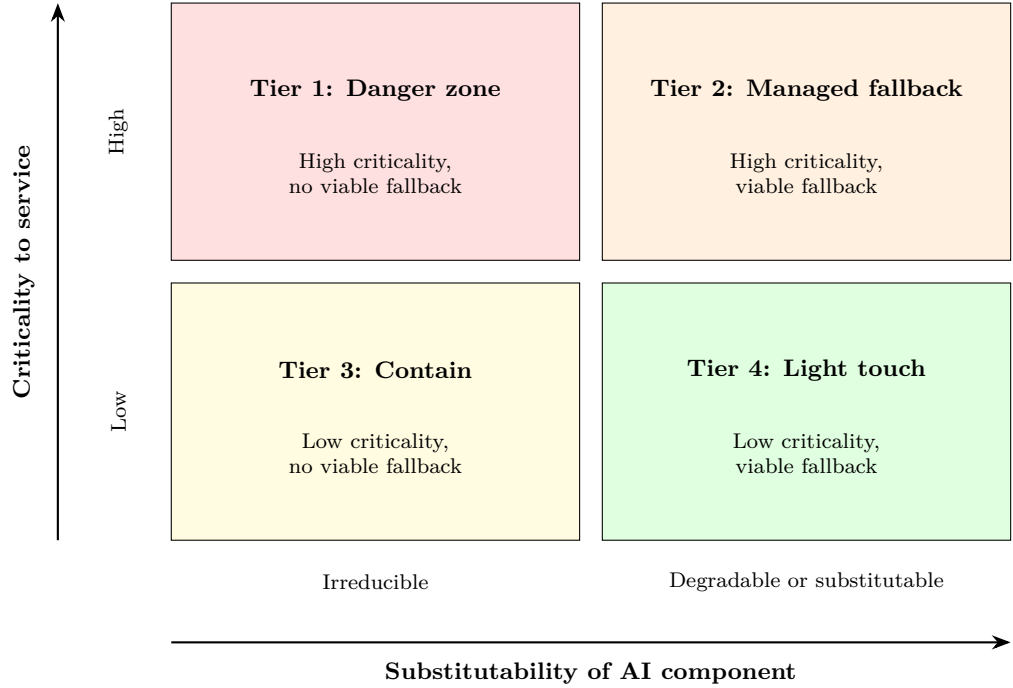
\begin{figure*}[tp]
\centering
\begin{tikzpicture}
    \fill[red!12]    (0,3.7)    rectangle (5.4,7.1);
    \draw[thin]      (0,3.7)    rectangle (5.4,7.1);
    \fill[orange!12] (5.7,3.7)  rectangle (11.1,7.1);
    \draw[thin]      (5.7,3.7)  rectangle (11.1,7.1);
    \fill[yellow!14] (0,0)      rectangle (5.4,3.4);
    \draw[thin]      (0,0)      rectangle (5.4,3.4);
    \fill[green!12]  (5.7,0)    rectangle (11.1,3.4);
    \draw[thin]      (5.7,0)    rectangle (11.1,3.4);

    \node[align=center, font=\small\bfseries] at (2.7,5.95) {Tier 1: Danger zone};
    \node[align=center, font=\footnotesize]   at (2.7,4.85) {High criticality,\\no viable fallback};
    \node[align=center, font=\small\bfseries] at (8.4,5.95) {Tier 2: Managed fallback};
    \node[align=center, font=\footnotesize]   at (8.4,4.85) {High criticality,\\viable fallback};
    \node[align=center, font=\small\bfseries] at (2.7,2.25) {Tier 3: Contain};
    \node[align=center, font=\footnotesize]   at (2.7,1.15) {Low criticality,\\no viable fallback};
    \node[align=center, font=\small\bfseries] at (8.4,2.25) {Tier 4: Light touch};
    \node[align=center, font=\footnotesize]   at (8.4,1.15) {Low criticality,\\viable fallback};

    \node[font=\footnotesize] at (2.7,-0.55)  {Irreducible};
    \node[font=\footnotesize] at (8.4,-0.55)  {Degradable or substitutable};
    \draw[-{Stealth[length=2.5mm]}, thick] (0,-1.35) -- (11.1,-1.35);
    \node[font=\small\bfseries] at (5.55,-1.75) {Substitutability of AI component};

    \node[rotate=90, font=\footnotesize] at (-0.7,5.4) {High};
    \node[rotate=90, font=\footnotesize] at (-0.7,1.7) {Low};
    \draw[-{Stealth[length=2.5mm]}, thick] (-1.5,0) -- (-1.5,7.1);
    \node[rotate=90, font=\small\bfseries] at (-1.95,3.55) {Criticality to service};
\end{tikzpicture}
\caption{The Criticality-Substitutability Matrix. Each mapped AI dependency is placed by its criticality to an important business service and the substitutability of the AI component, yielding four tiers of resilience treatment.}
\label{fig:matrix}
\end{figure*}

Figure~\ref{fig:matrix} presents the resulting matrix. Combining a binary reading of criticality with the substitutability classes yields four practical tiers, summarised in Table~\ref{tab:tiers}. The tier a dependency occupies determines the depth of resilience treatment it warrants, allowing firms to concentrate effort where the exposure is greatest rather than applying uniform controls to every AI component.

\begin{table*}[tp]
\centering
\caption{Resilience treatment by tier under the Criticality-Substitutability Matrix.}
\label{tab:tiers}
\small
\begin{tabular}{L{2.0cm} L{3.2cm} L{8.0cm}}
\toprule
\textbf{Tier} & \textbf{Profile} & \textbf{Resilience treatment} \\
\midrule
Tier 1 & High criticality, irreducible & The danger zone. Requires provider-level resilience assurance, exit and continuity planning, explicit board visibility, and treatment as a candidate concentration exposure. Firms should actively work to move these dependencies out of Tier 1 by engineering substitutability. \\
\addlinespace
Tier 2 & High criticality, degradable or substitutable & Manage and rehearse the fallback. Define the impact tolerance for operation with the AI component degraded or removed, and test reversion regularly so the fallback remains real. \\
\addlinespace
Tier 3 & Low criticality, irreducible & Contain and monitor. The exposure is bounded by low criticality, but scope creep can raise criticality over time, so guard against silent promotion into critical services. \\
\addlinespace
Tier 4 & Low criticality, substitutable & Light-touch. Standard monitoring and change control are proportionate; resilience effort is better spent on higher tiers. \\
\bottomrule
\end{tabular}
\end{table*}

\subsection{Step Three: Extend Impact Tolerances to AI-Specific Failure Modes}

An impact tolerance expressed only as a tolerable duration of unavailability is insufficient for an AI dependency, because, as Section 3 argued, the characteristic AI failure is degradation rather than outage. Firms should therefore extend the impact tolerance for AI-dependent services to encompass correctness as well as availability, and define the point at which a degrading model is treated as failed for tolerance purposes. This requires instrumentation the resilience regime does not conventionally demand: monitoring for distributional drift, validation of outputs against challenge or canary sets, and thresholds that trigger fallback when correctness falls below an acceptable level. The practical test is simple to state and demanding to meet. A firm should be able to detect that an AI-dependent service has begun to fail before its customers or its regulator do, and it should have defined in advance how bad is too bad.

\subsection{Step Four: The Fallback Doctrine}

The fourth step is the most consequential and the most frequently neglected. For any dependency above Tier 4, resilience depends on a fallback: a path by which the service continues, in degraded or manual form, when the AI component is lost. The central claim of this framework is that a fallback that is documented but not resourced is not a control. Human oversight, in particular, is routinely cited as the mitigation for AI failure, yet human oversight is a resilience control only if the human path is staffed, retains the competence to operate without the AI, and is exercised often enough to remain viable. Where an AI system has replaced a manual process, and that process has been decommissioned or allowed to atrophy, the fallback is fictional, and the dependency is in truth irreducible whatever the register claims. Firms should test their fallbacks by exercising them, should retain the capability that fallbacks assume, and should treat the discovery of a fictional fallback as a resilience finding of the first order. For irreducible dependencies that cannot be given a fallback, the appropriate response is exit planning and, where feasible, architectural work to raise substitutability, for example by abstracting model access behind an interface that permits a provider to be replaced.

\subsection{Step Five: Provider-Level Concentration Management}

The final step lifts the analysis from the individual dependency to the provider. Foundation-model providers should be brought into the firm's third-party risk management on the same footing as other critical suppliers, and assessed not only for the assurance they can offer but for the firm's own ability to substitute them. Where a provider underpins several services, its aggregate criticality may exceed that of any single dependency, and it should be assessed as a concentration exposure in its own right, with reference to the Critical Third Parties regime and, for firms in scope, the Digital Operational Resilience Act oversight framework \cite{ss624ctp2024, dora2022}. Architecturally, substitutability at the provider level is advanced by abstraction layers that decouple applications from any single model interface, by avoiding provider-specific features where portability matters, and by maintaining the practical ability to fail over to an alternative. The Bank of England's work toward a shared-responsibility model for AI deployment is directly relevant here, since it clarifies which resilience obligations rest with the provider and which remain with the firm \cite{boefpcai2025}.

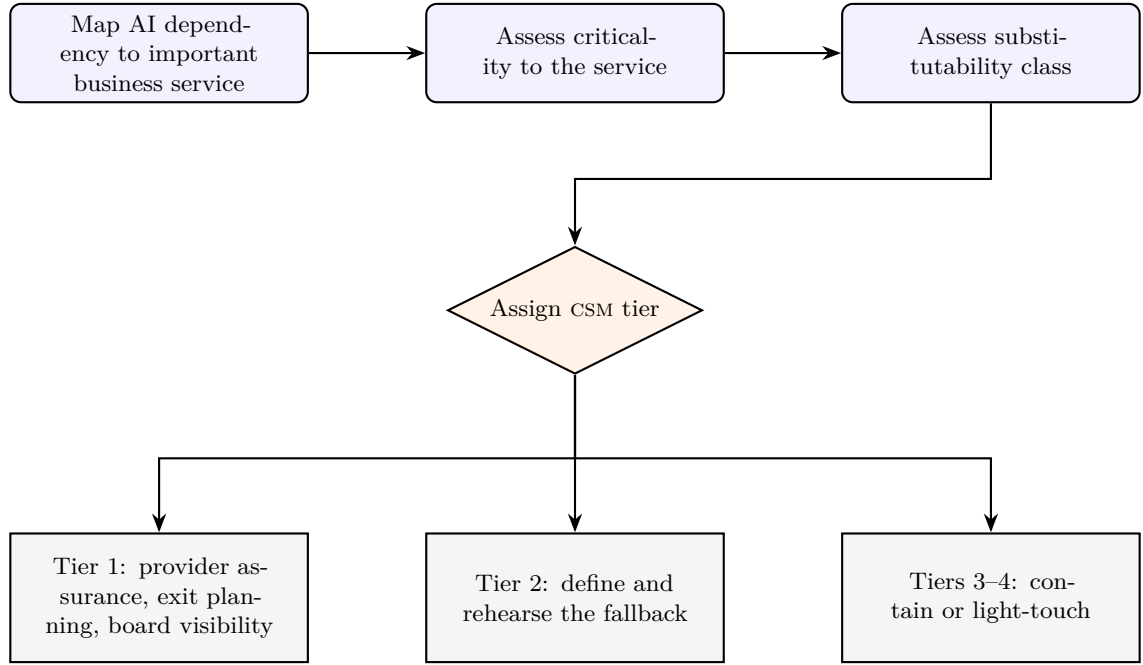
\begin{figure*}[tp]
\centering
\begin{tikzpicture}
    \node[proc, text width=3.7cm, minimum height=1.3cm] (map)  at (2.6,0)  {Map AI dependency to important business service};
    \node[proc, text width=3.7cm, minimum height=1.3cm] (crit) at (8.1,0)  {Assess criticality to the service};
    \node[proc, text width=3.7cm, minimum height=1.3cm] (subs) at (13.6,0) {Assess substitutability class};

    \node[dec, text width=2.6cm, minimum height=1.5cm] (tier) at (8.1,-3.4) {Assign \CSM tier};

    \node[term, text width=3.7cm, minimum height=1.7cm] (t1)  at (2.6,-7.2)  {Tier 1: provider assurance, exit planning, board visibility};
    \node[term, text width=3.7cm, minimum height=1.7cm] (t2)  at (8.1,-7.2)  {Tier 2: define and rehearse the fallback};
    \node[term, text width=3.7cm, minimum height=1.7cm] (t34) at (13.6,-7.2) {Tiers 3--4: contain or light-touch};

    \draw[flow] (map) -- (crit);
    \draw[flow] (crit) -- (subs);
    \draw[flow] (subs.south) -- ++(0,-1.0) -| (tier.north);
    \draw[flow] (tier.south) -- (t2.north);
    \draw[flow] (tier.south) -- ++(0,-1.1) -| (t1.north);
    \draw[flow] (tier.south) -- ++(0,-1.1) -| (t34.north);
\end{tikzpicture}
\caption{The AI Resilience Framework as a decision workflow. Dependencies are mapped, assessed on both axes, assigned a tier, and given proportionate resilience treatment.}
\label{fig:workflow}
\end{figure*}

Figure~\ref{fig:workflow} summarises the framework as a decision workflow from dependency discovery to tiered treatment.

\subsection{Where the Framework Sits}

The framework deliberately locates AI resilience within the operational resilience function and its accountable senior manager rather than within an AI ethics or model-risk committee. Model risk and resilience are related but distinct, and the connection between them, that drift is both a model-risk signal and a resilience signal, is a reason to join the two registers, not to subordinate resilience to model governance. Firms that maintain a single obligation register spanning trustworthy-AI and resilience requirements will avoid the duplication and the blind spots that the current bifurcation produces, in keeping with the convergence principle established for the broader regulatory stack \cite{shelby2026regstack}.

\subsection{Mapping the Framework to Regime Obligations}

The framework is regime-agnostic by design, but its outputs are not regulatory orphans. Each step produces evidence that discharges obligations across several instruments at once, which is the practical expression of the convergence principle: one well-constructed body of work, many satisfied requirements. Table~\ref{tab:mapping} sets out the mapping. The point is not that the framework replaces any regime's own requirements, but that a firm executing it generates, as a by-product, much of the evidence those regimes demand, and does so from a single register rather than several.

\begin{table*}[tp]
\centering
\caption{Mapping AI Resilience Framework outputs to regime obligations.}
\label{tab:mapping}
\small
\begin{tabular}{L{3.2cm} L{5.6cm} L{6.4cm}}
\toprule
\textbf{Framework step} & \textbf{Output} & \textbf{Obligations supported} \\
\midrule
Step 1: Dependency mapping & Register of AI dependencies indexed by important business service & UK operational resilience mapping duty; DORA ICT asset and dependency inventory; EU AI Act system inventory \\
\addlinespace
Step 2: Criticality-substitutability tiering & Tiered classification of every AI dependency & UK operational resilience prioritisation; DORA criticality assessment; risk-based allocation of effort under all regimes \\
\addlinespace
Step 3: Extended impact tolerances & Tolerances covering correctness as well as availability, with degradation thresholds & UK operational resilience impact tolerances; DORA resilience testing; model performance monitoring under SS1/23 \\
\addlinespace
Step 4: Fallback doctrine & Tested reversion and continuity paths; identification of fictional fallbacks & UK operational resilience severe-but-plausible testing; DORA business continuity; human-oversight expectations under the EU AI Act \\
\addlinespace
Step 5: Provider concentration management & Provider-level exposure and substitution assessment & UK Critical Third Parties regime; DORA third-party and concentration risk; EU AI Act provider obligations mapped to deployer reliance \\
\bottomrule
\end{tabular}
\end{table*}

\section{Worked Application}

To make the framework concrete, consider three illustrative AI dependencies of the kind found in a mid-to-large United Kingdom financial services firm. The examples are stylised, but the classifications and treatments follow directly from the framework.

\paragraph{A generative-AI customer-service assistant.} A firm deploys a large-language-model assistant to handle first-line customer queries within a customer-servicing important business service. The assistant is highly visible and contributes materially to the service, so its criticality is high. However, the firm retains a staffed contact centre capable of handling the volume, so the dependency is degradable, placing it in Tier 2. The correct treatment is to define an impact tolerance for operation with the assistant unavailable or degraded, to ensure the contact centre retains the capacity and competence to absorb the load, and to rehearse the reversion. The resilience risk here is not the assistant's failure but the temptation to reduce the human contact centre below the level the fallback requires, which would silently move the dependency toward Tier 1.

\paragraph{An AI transaction-monitoring model.} A firm replaces a deterministic rules engine for transaction monitoring with a machine-learning model embedded in a payments important business service. Its criticality is high. If the legacy rules engine has been decommissioned and no equivalent non-AI path remains, the dependency is irreducible and sits in Tier 1. This is the danger zone, and the framework directs the firm to obtain provider and model assurance, to instrument for silent degradation, since a monitoring model that quietly drifts will fail open or closed without an outage, and to undertake exit or continuity planning, potentially including the retention of a maintained fallback ruleset. The classification exposes a decision that model-risk governance alone would not surface: the firm has accepted an irreducible dependency in a critical service, and it must either accept and manage that exposure explicitly or invest to make it degradable.

\paragraph{A shared frontier-model provider.} A single external frontier-model provider underpins the customer-service assistant, several internal productivity tools, and a document-analysis capability used in onboarding. No individual use may be critical, but in aggregate the provider supports functions across multiple services, and its loss or degradation would be felt simultaneously across all of them. Assessed at the provider level under Step Five, this is a concentration exposure that warrants treatment as a candidate critical third party, with sector-level correlation in view and architectural work to preserve the ability to substitute the provider. This is the exposure that firm-level model governance is structurally unable to see, because it manifests only when the individual dependencies are aggregated.

\section{Implications for Practitioners}

The framework has distinct consequences for each function that touches AI in a regulated firm.

\emph{For chief information security officers and accountable senior managers}, the immediate task is to bring AI dependencies into the resilience register and to take ownership of the fallback doctrine. The accountability that the Senior Managers and Certification Regime attaches to important business services extends to the AI components within them, and the joint statement of May 2026 makes explicit that boards and senior management are expected to understand frontier AI risk \cite{boefcahmtfrontier2026}. A senior manager who cannot describe the fallback for a critical AI dependency has an accountability gap, not merely a technical one.

\emph{For security architects}, the framework translates into design obligations: abstract model access so that providers can be substituted, instrument services for grey failure rather than only for outage, and contain autonomous components so that their errors cannot propagate beyond a defined boundary. The identity and containment disciplines developed for Zero Trust architectures apply directly to AI components, which are workloads that require scoped, short-lived credentials and constrained egress like any other. Designing for substitutability is the single most valuable architectural contribution to AI resilience, because it is what moves a dependency out of Tier 1.

\emph{For model-risk and data functions}, the framework reframes drift monitoring as a resilience control as well as a model-risk control, and connects the model risk management expectations of SS1/23 to the operational resilience regime \cite{ss123mrm2023}. The measurement infrastructure that model-risk teams build to detect performance degradation is precisely the instrumentation that Step Three requires, and sharing it across the two registers avoids duplicated effort.

\emph{For boards and compliance functions}, the framework offers a single lens across obligations that currently arrive from different directions. A firm can map its AI dependencies once and use that map to satisfy the documentation and oversight expectations of the trustworthy-AI stack, the impact-tolerance and substitution expectations of the resilience regime, and the concentration expectations of the Critical Third Parties framework, rather than maintaining a separate assessment for each. This is the AI-specific instance of the convergence argument: the obligations overlap on the same underlying dependencies, and a single well-constructed register can serve them all.

\section{Limitations and Future Work}

Several limitations qualify the framework and mark directions for further work. First, the criticality-substitutability classification is presented as a qualitative instrument, and its two axes admit finer measurement than the binary and three-class readings used here. A natural extension is a quantitative substitutability score, drawing on the resource-constrained metric tradition developed elsewhere in this body of work \cite{shelby2026regstack}, that would express the effort and time required to substitute a dependency and allow tiering thresholds to be calibrated rather than asserted. Second, the framework treats provider concentration primarily from the perspective of the individual firm, whereas the systemic dimension, in which many firms share a dependency invisible to each, can only be assessed with sector-level visibility of the kind that the Critical Third Parties regime and the Financial Policy Committee are positioned to supply. Formalising the relationship between firm-level substitution assessments and sector-level correlation is an open problem with direct policy relevance.

Third, the treatment of AI-specific failure modes in Step Three assumes that firms can instrument for drift and degradation, which is achievable for models a firm hosts or can observe closely, but is considerably harder for capability consumed through an external interface whose internals are opaque and whose version may change without notice. The shared-responsibility model under development by United Kingdom authorities \cite{boefpcai2025} is the natural venue for resolving where that monitoring obligation should rest. Fourth, the framework concentrates on AI as a dependency and treats AI as a threat only in outline; the interaction between the two, in which an AI-enabled attack degrades an AI-dependent service, is an emerging area that the May 2026 joint statement began to address \cite{boefcahmtfrontier2026} and that warrants dedicated treatment. Finally, the framework has been developed for the United Kingdom and European context; its transferability to other regulatory regimes, and its interaction with the extraterritorial reach of the EU AI Act for firms operating across jurisdictions, remains to be examined.

\section{Conclusion}

The governance response to artificial intelligence in regulated firms has been substantial, but it has been oriented toward trustworthiness, and trustworthiness is not resilience. A firm can hold an AI system that is safe, fair, documented, and overseen, and that is at the same time an irreducible, unsubstitutable dependency embedded in a service the firm is obliged to keep running. That is the AI resilience gap, and it is where the material risk of AI adoption now concentrates.

United Kingdom authorities are closing the gap from their side. The Financial Policy Committee has identified AI concentration as a systemic concern, the Critical Third Parties regime is poised to bring major AI and cloud providers within supervisory reach, and the May 2026 joint statement has reframed frontier AI as a resilience threat as well as a governance subject. The Financial Conduct Authority has been clear that AI-specific rules are not coming, which means the obligation to make AI resilient will be discharged through the existing resilience regime rather than a bespoke one. Firms that wait for prescriptive AI rules will find that the applicable rules were already in force.

The AI Resilience Framework offers a route to discharge that obligation. By mapping AI dependencies to important business services, classifying them on the criticality-substitutability matrix, extending impact tolerances to the degradation failures that characterise AI, enforcing a fallback doctrine that distinguishes real fallbacks from fictional ones, and managing provider concentration explicitly, a firm can move from an AI governance policy, which describes intentions, to demonstrable resilience, which survives disruption. Resilience for AI, like security for the wider estate, is an architecture and operating-model problem, not a policy-document problem. The firms that internalise this now, while the supervisory perimeter is still forming, will be answering the next examination from a position of knowledge rather than reconstructing their dependencies during the incident that exposes them.

\bibliography{shelby-master}

\end{document}